\documentclass[journal=nalefd,manuscript=letter,layout=traditional]{achemso}
\usepackage{latexsym}
\usepackage{amssymb}
\usepackage{graphicx}
\usepackage{dcolumn}
\usepackage{sidecap}
\usepackage{bm}
\usepackage{graphicx}
\usepackage{hyperref}
\setkeys{acs}{keywords = true}

\author{Vitaly P. Panov}
\affiliation{Department of Electronic and Electrical Engineering,
	Trinity College, University of Dublin, Dublin 2, Ireland}

\author{Sithara P. Sreenilayam}
\affiliation{Department of Electronic and Electrical Engineering,
	Trinity College, University of Dublin, Dublin 2, Ireland}

\author{Yuri P. Panarin}
\affiliation{Department of Electronic and Electrical Engineering,
	Trinity College, University of Dublin, Dublin 2, Ireland}
\alsoaffiliation{School of Electrical and Electronic Engineering, Dublin Institute of Technology, Dublin 8, Ireland}

\author{Jagdish K. Vij}
\email{jvij@tcd.ie}
\affiliation{Department of Electronic and Electrical Engineering,
	Trinity College, University of Dublin, Dublin 2, Ireland}

\author{Chris J. Welch}
\affiliation{Department of Chemistry, University of Hull, HU6 7RX, UK}

\author{Georg H. Mehl}
\affiliation{Department of Chemistry, University of Hull, HU6 7RX, UK}

\title{Characterization of the sub-micrometer hierarchy levels in the twist-bend nematic phase with nanometric helices via photopolymerization. Explanation for the sign reversal in the polar response.}

\keywords{twist-bend nematic; hierarchical self-assembly; polymerization; Scanning Electron Microscopy}

\begin{document}

\begin{tocentry}
\centerline{\includegraphics[width=2.2 in]{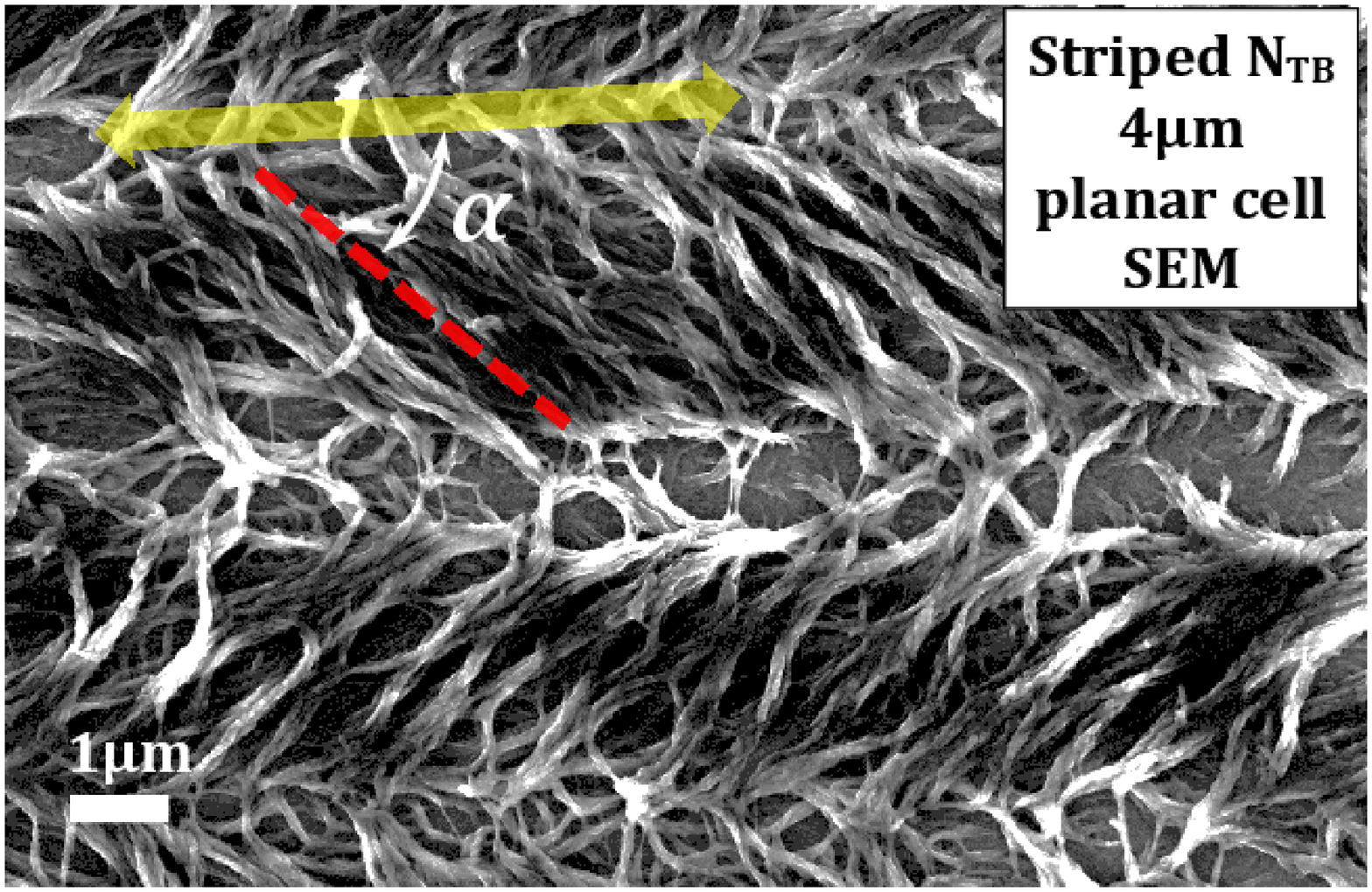}}
\end{tocentry}

\begin{abstract}
Photo-polymerization of a reactive mesogen mixed with a mesogenic dimer, shown to exhibit the twist-bend nematic phase ($N_{TB}$), reveals the complex structure of the self-deformation patterns observed in planar cells. The polymerized reactive mesogen retains the structure formed by liquid crystalline molecules in the twist bend phase, thus enabling observation by Scanning Electron Microscope (SEM). Hierarchical ordering scales from tens of nanometers to micrometers are imaged in detail. Submicron features, anticipated from earlier X-ray experiments, are visualized directly. In the self-deformation stripes formed in the $N_{TB}$ phase, the average director field is found tilted in the cell plane by an angle of up to 45$^{\circ}$ from the cell rubbing direction. This tilting explains the sign inversion being observed in the electro-optical studies.
\end{abstract}



\vspace{5mm}

Liquid crystals (LCs) are known to form complex spatial structures that scale from the molecular length via nanometric assemblies such as in the Sm$C_{\alpha}$*, blue phase, dark conglomerate phase or twist-grain-boundary phase to the sample size (cholesteric phase, $N_{TB}$) \cite{Handbook}. Beyond stimulating major theoretical advances in topology\cite{2017Smalyukh} and mean field theory\cite{2013SelingerPRE}, LCs have made the stunning success of current information display and photonic devices industries possible and are at the forefront of emerging technologies relating to light and matter. The major basis for applications of LCs in photonic devices so far is the three-dimensional deformation of the LC director on the application of a relatively weak external field. Meanwhile, preserving the molecular structure of a phase by polymerisation will not only extend the range of potential applications, but will enable further structural studies by methods that are normally unsuitable for liquid samples\cite{DierkingRSC3013}. Synthesis of helical polyacetylene for potential nano-sized polymer solenoids is an example\cite{2009Akagi}.

Reactive mesogens or polymerizable LCs are designed to have both LC properties (including miscibility with other liquid crystalline phases/materials) and polymerisation controlled by light, heat, concentration of oxygen, etc. This class of materials have been shown to aid in the liquid crystalline alignment \cite{2011APLKim,2000APLKang,Varanytsia,ZumerBook} and may also extend the temperature range of the mesophases\cite{2015HelenAPL}.

The twist-bend nematic phase has recently become one of the most topical areas of research in the field of LCs due to the presence of a complicated hierarchy of periodic structures\cite{Hierarchy}. The phase is observed in wide range of materials and mixtures containing either odd-hydrocarbon-chain-link dimers or bent-core mesogenic materials. Mixing of the materials provides freedom for control of operating temperature range, that includes room temperature. This, combined with a relatively high tolerance to mixture compositions (up to 40$\%$ of added non-$N_{TB}$ components), makes the phase "ready to use" for potential applications\cite{2011TripathiPRE,2014MGTPMat}. A combination of the flexoelectric properties and nanometer-scaled helical pitch\cite{APL11,2013LuckhurstPRL} provides one of the fastest electro-optic responses (a few microseconds) observed so far in LCs. Though yet not optimized for devices (the switching angle is very low), it demonstrates in principle the potential of making use of emerging chirality for photonic applications; and, beyond that, for chiral synthesis\cite{APL12,BCI} of pharmaceutical drugs.

The scale of the structures in the $N_{TB}$ phase ranges from the sample size/thickness, i.e. several micrometers to several nanometers. Some features are found to be of the submicron size\cite{Hierarchy}. Meanwhile, optical microscopy, though extremely useful, is restricted to have its resolution limited by optical wavelength. In this Letter, we demonstrate that the technique of photopolymerization is extremely effective for deciphering the complex hierarchical structures in the $N_{TB}$ phase using SEM.

The structural formulae of the materials used in the experiment are shown in Figure 1. Mixture of mesogenic dimer CB-C7-CB (approximately 70 w/w $\%$ of the total mixture) with 4-Cyano-4'-pentylbiphenyl ($\sim$17.7 w/w\% of 5CB in the total mixture) was prepared to lower the phase transitions temperatures and avoid thermally induced polymerization. The compound 1,4-Bis-[4-(3-acryloyloxypropyloxy)benzoyloxy]-2-methylbenzene (purchased from Synthon Chemicals, cat. No. ST03021), known as a reactive mesogen RM 257, was used as a polymerizable component of the mixture ($\sim$12 w/w\%). Rod-shaped, forming a wide nematic phase, molecules of RM 257 provide good miscibility and retain the structure after polymerization. The photoinitiator Irgacure$^{\textregistered}$ 819 ($\sim$0.3 w/w\% of the total mixture) was chosen to ensure rapid polymerization by radiation with commercially available ultraviolet light emitting diodes (LEDs). The four-component mixture possessed an isotropic to nematic phase transition at $\sim105^{\circ}C$ and the nematic to $N_{TB}$ transition at $\sim55^{\circ}C$ on cooling.

\begin{figure}[htbp]
\centerline{\includegraphics[width=3.35 in]{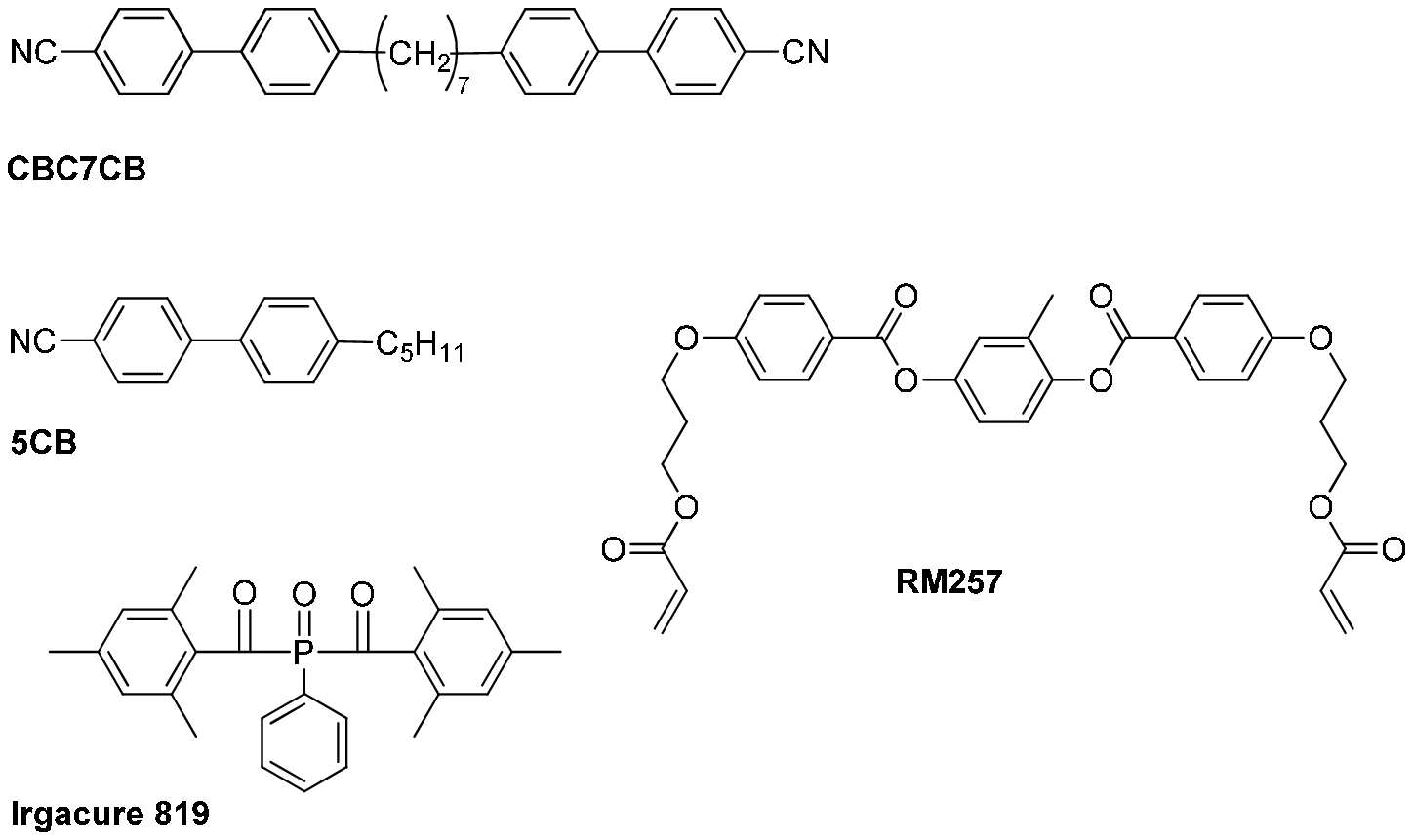}}
\label{fig1}\caption{Chemical formulae of materials used in the experiment.}
\end{figure}

The mixture containing the LCs, the reactive mesogen and the photoinitiator was filled in commercial sandwich cells (KSXX-0X/XX11P6NSS05 EHC$^{\textregistered}$, Japan) by capillary action at 95$^{\circ}C$ and was immediately cooled to 35-55$^{\circ}C$ in the $N_{TB}$ phase to avoid temperature-induced polymerization. A microscope-mounted hot-stage connected to a Eurotherm$^{\textregistered}$ 2604 temperature controller was used to stabilize the temperature within $\pm$0.1$^{\circ}C$. The textures were observed using polarising optical microscope (POM) Olympus$^{\textregistered}$ BX51 equipped with 5X, 10X or 20X objective lens and a filter used for blocking the UV radiation of the light source. To induce photopolymerization, an ultraviolet LED with peak emission wavelength at 390 nm was mounted onto the camera port of the microscope, thus UV light was projected onto a half-millimeter diameter spot on the sample plane. The optimal exposure time of the order of 20 seconds could be controlled by varying concentration of the photo-initiator. This set-up enabled us to preserve the structures by UV polymerization for several experimental conditions of interest (e.g. at different temperatures) for each sample cell. The polarizer of the microscope was removed from the path of the UV light during exposure. It is noted that introduction of a prism polarizer did not show any visible influence on the polymerized structures indicating that the absorption of UV by the sample was practically isotropic.

Heating the sample to a higher temperature phase can be used to evaluate results of polymerization process as texture in the irradiated area is preserved and shows a clear contrast to the surroundings (see supplementary information). When the sample was heated to the isotropic state, areas of the cell polymerized in LC phases, still exhibited birefringence when the cell was observed between the crossed-polarizers. On the other hand, areas polymerized in the isotropic state remained isotropic (i.e. dark texture observed between the crossed polarizers) on consequent cooling to LC phases.

 When the polymerization was complete, two glass plates of the cell are pulled apart and the liquid crystalline material was washed away with acetone. The polymer "sponge" remained attached to the glass surfaces and could be used for investigating the preserved structures. Although the polymerized RM257 has low conductivity, using thin samples (up to 4$\mu$m) and the Indium Tin Oxide (ITO)-coated glass plate as the ground electrode enabled imaging by the state-of-the-art scanning electron microscope (SEM) (Carl Zeiss Ultra Plus$^{\textregistered}$) with a resolution of  approximately 20 nm.

\begin{figure*}[htbp]
\centerline{\includegraphics[width=6.9in]{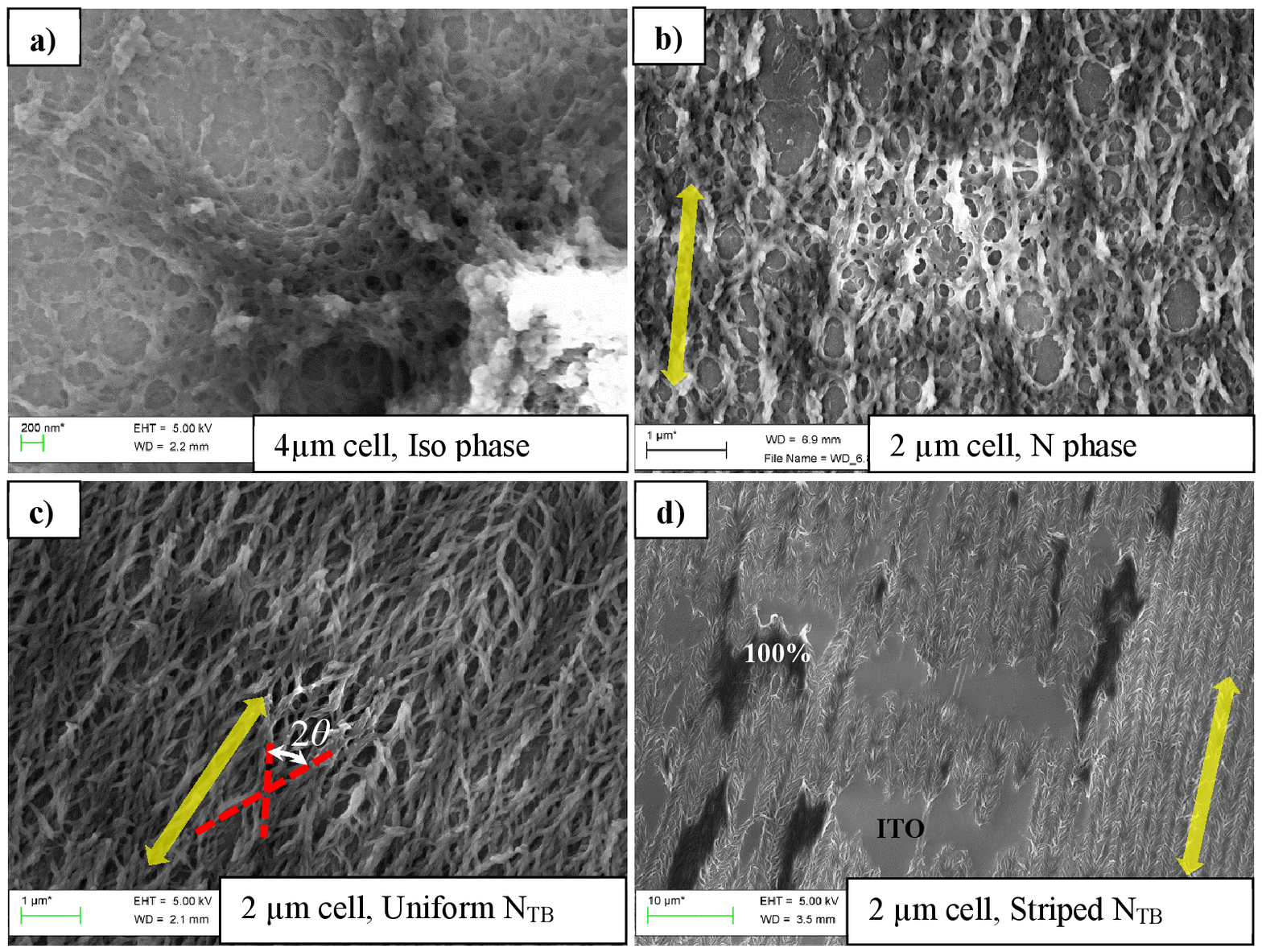}}
\label{fig2}\caption{SEM images of planar sample polymerized in different phases. a) Isotropic phase, b) Nematic phase, c) Higher temperature uniform pattern in $N_{TB}$ phase, d) Striped pattern in $N_{TB}$ phase.}
\end{figure*}
Figure 2 shows SEM images of a planar-aligned sample, polymerized at different temperatures. Differences between the isotropic (2a), nematic (2b) and $N_{TB}$ (2c, d) phases are clearly observed. The isotropic (Figure 2a) phase is characterized by the lack of a preferred orientation. The nematic phase (Figure 2b) shows a low-ordered structure; the cell rubbing direction (shown by the yellow arrow) being clearly observed. The $N_{TB}$ phase exhibits a range of structures that depend on the temperature of polymerisation of the sample and its confinement conditions. Although in the vicinity of the $N$-$N_{TB}$ phase transition POM textures of both phases are practically uniform, a detailed SEM image clearly shows a difference as the average direction (red lines) of the polymer filaments governed by the $N_{TB}$ phase (Figure 2c) form an angle $\alpha$ with the rubbing direction (yellow arrow). The most spectacular patterns are observed in the lower temperature region of the $N_{TB}$ phase (Figure 2d) where the characteristic self-deformation stripes are clearly visible. In Figure 2d one observes the polymer "sponge" splitting between the two glass plates (the top and the bottom substrates of the cell) in three major ways: (i) all of the polymer remains on the imaged glass plate (marked as 100\%), (ii) all of the polymer is attached to the second glass plate leaving the ITO surface bare (marked as ITO), and, in most cases, (iii) the polymer structures split somewhat equally between the two substrates.

Dimensions of the hierarchical structures were found to range from as low as a 8-12 nm helix\cite{2013LavrNatComm,2013ClarkPNAS,2017StevensonPCCP} to sizes of the pattern comparable to the thickness of the cell\cite{PanovPRL,APL12}, thus spanning over 3 orders of magnitude in size. Figure 3 shows a series of SEM micrographs depicting the characteristic features of the planar-aligned cells under varying magnification.

\begin{figure*}[htbp]
\centerline{\includegraphics[width=6.9in]{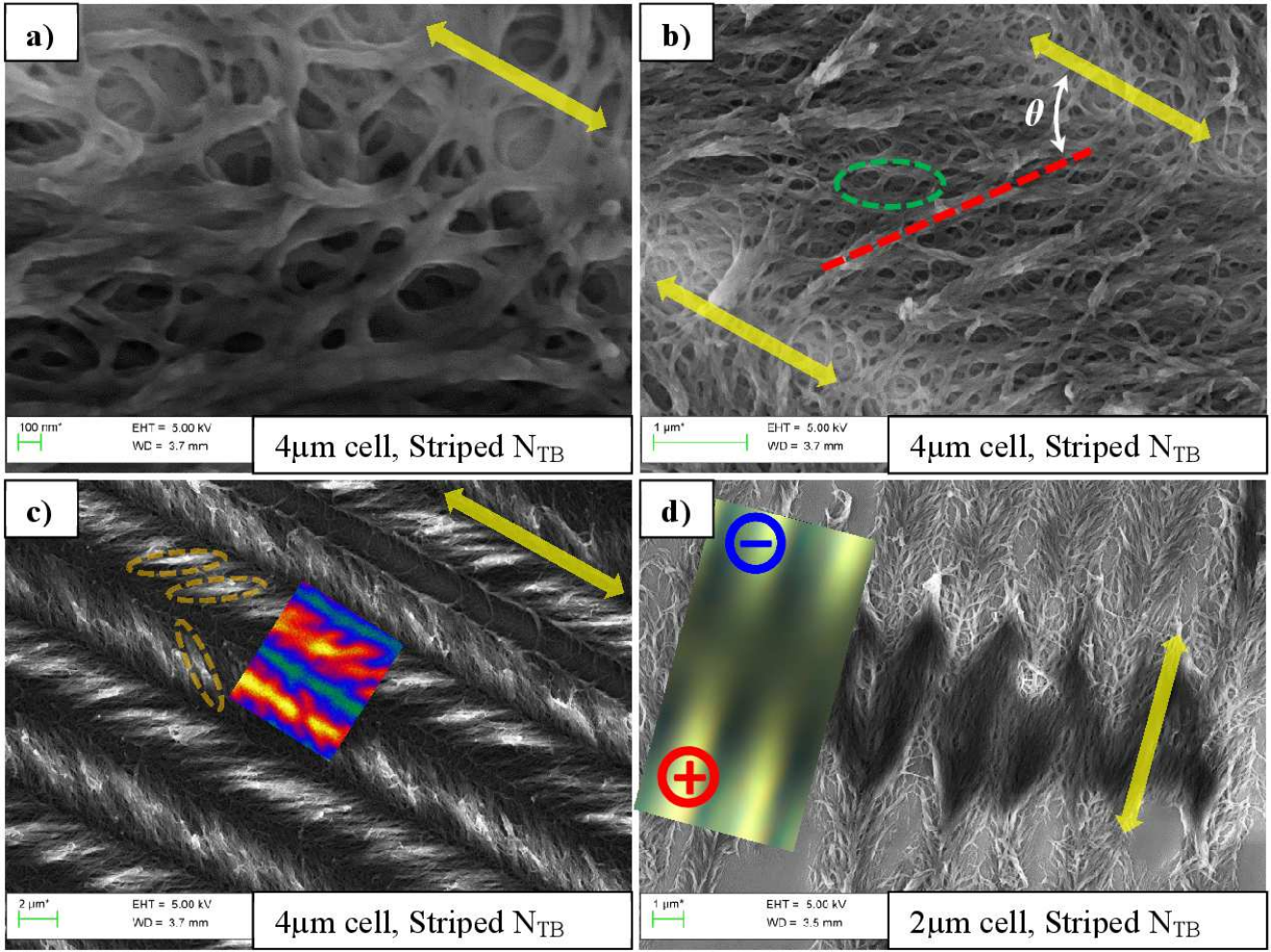}}
\label{fig3}\caption{Hierarchy of the polymerized $N_{TB}$ patterns in planar cells. Yellow arrow represents cell rubbing direction. a) High-magnification SEM image of polymer filaments. b) Sub-micrometre scale pattern. Part of a self-deformation stripe is visible. Average fibre direction (red dashed line) forms a large angle $\alpha$ with the rubbing direction. Green oval highlights submicron helical patterns c) Self-deformation stripes in thicker cells. Elements providing impression of "rope-like textures" are highlighted by orange ovals. Color insert - corresponding FCPM image scaled from Figure 2c in \cite{2013SPIE}. d) Possible boundary between domains with opposite handedness. Most of the polymer adheres to one glass plate. Color insert - fragment of corresponding POM texture scaled from Figure 3a in\cite{APL11}.}
\end{figure*}
The polymer filaments have a finite thickness, and, therefore, impose a lower limit on the resolution of the technique. Figure 3a shows that the filaments as thin as 20 nm can be observed. Thus the resolution of our experiment is not sufficient to give a direct confirmation of the existence of a duplex double-helical DNA-like chains proposed from resonant X-ray studies\cite{ClarkDH,GoreckaDH}. However, it is reasonably possible to envisage that such chains may exist in the material. Moreover, Figures 3a, b show that axes of such chains (or other nm-scale units) are not straight and, in their turn, form helical patterns on the scale of 100-200 nm. Though it is difficult to detect more than a couple of undistorted periods (highlighted by green oval in Figure 3b) at this scale due to a relatively low order and the influence of larger-scale patterns. These are likely to correspond to the $\sim0.1$ $\mu$m correlation length reported from X-ray experiments\cite{2017StevensonPCCP}.

At the sample scale, the average direction of the polymer fibres forms an angle $\alpha$ with the rubbing direction (Figure 3b, 3c, 3d). Once the stripes are formed on cooling, $\alpha$ reaches saturated values of 40-45$^{\circ}$. Stripes of double cell gap periodicity are formed with neighboring bands tilted with an opposite sign of $\alpha$ (Figure 2d, 3c, 3d). Thus the self-deformation periodic patterns\cite{PanovPRL} can  be easily identified and compared with  textures observed by other methods.

Figures 3b and 3c provide details of the apparent "rope-like textures" visible under certain conditions in thicker cells by POM\cite{2011LuckhurstPRE,2017DenaLC} and the Fluorescent Confocal Polarizing Microscopy (FCPM)\cite{2013SPIE}. We note that the filaments form "fish-bone" patterns with submicron to micron periodicity (Figure 3c), caused most likely by intermodulation of the helical patterns with finer (100-200 nm) periodicities (Figure 3b). A helical structure with a micrometer-scale pitch is not being observed, as would be expected from the "rope-like" patterns appearing in optical polarizing  microscopy. Moreover, should the helix be present, the "fish-bone" patterns viewed in the optical range would change the direction when the microscope is refocused from one glass plate to the second. This is not observed.

\begin{figure}[htbp]
\centerline{\includegraphics[width=3.2in]{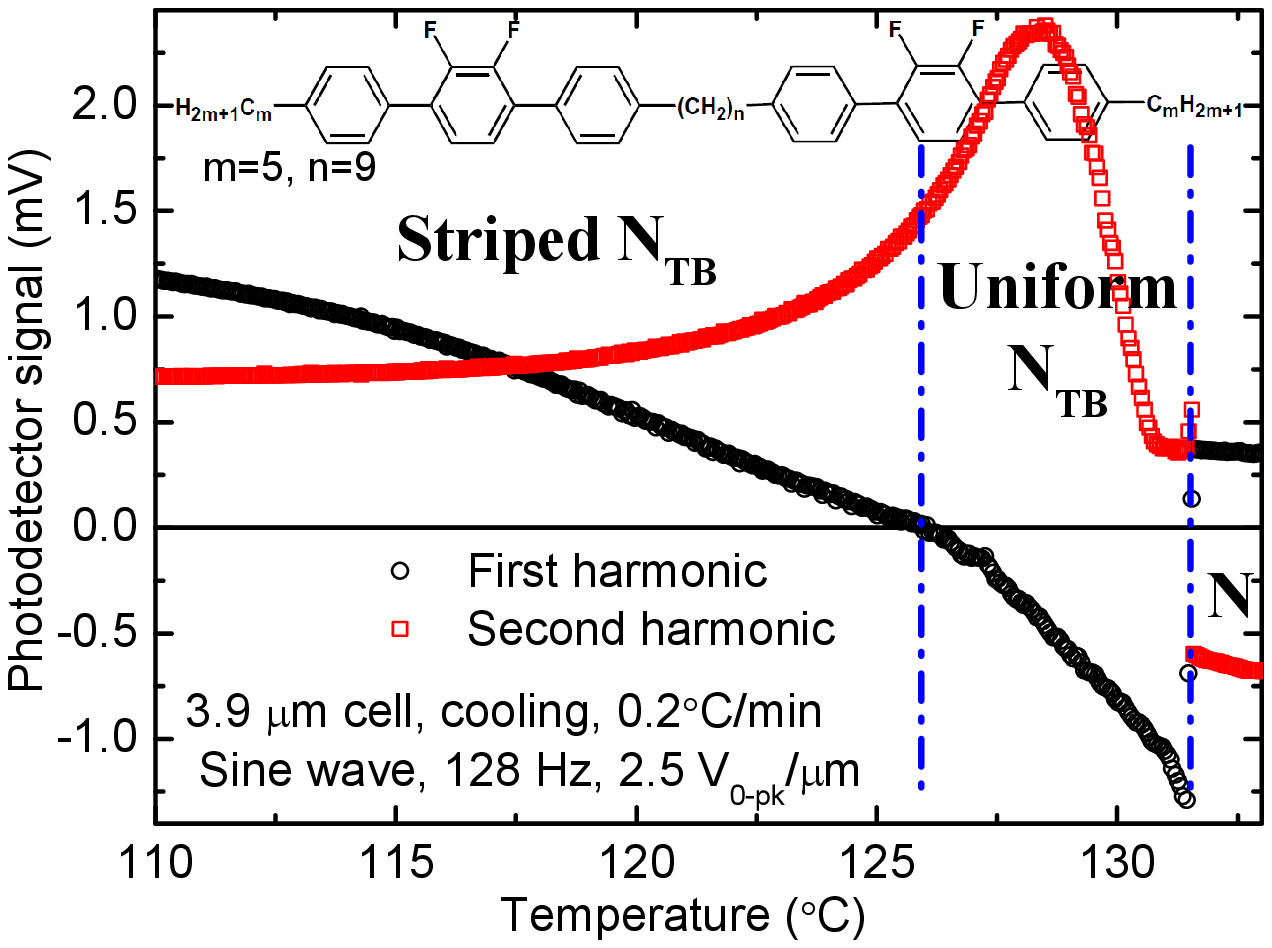}}
\label{fig4}\caption{Temperature dependence of the electro-optic response of DTC5C9 (chemical structure shown). Sign inversion of the first harmonic signal is visible in the temperature region corresponding to the onset of the stripes.}
\end{figure}
As reported earlier\cite{APL11}, an application of the electric field reveals rather large (up to sample size) structurally chiral domains switching in the opposite directions at a speed corresponding to the deformation of the nanometer-scale helix\cite{2013LuckhurstPRL}. The boundaries between such domains can be identified in POM images by characteristic discontinuities in the striped patterns present even in the absence of an electric field (see Figure 3a,b in\cite{APL11}). In SEM images such boundary areas are also identifiable by a reduced value of $\alpha$. The polymer "sponge" in the area tends to stay on one of the glass plates rather than being split. These observations will help in analysing a variety of domain boundary scenarios discussed by C. Meyer \textit{et. al.} (Figure 7 in \cite{2015Meyer}). However, for complete understanding of the domain boundary formation, information about all levels of hierarchy is required. Knowledge of the local heliconical angle of the nano-scale pitch is needed. This angle may differ from the fish-bone angle $\alpha$ being discussed here.

Far away from the domain boundary areas, the angle $\alpha$ is onset to an initial value of 25-30$^{\circ}$ (Figure 2c) at the N to $N_{TB}$ phase transition and then increases rapidly on cooling to a saturated value of 40-45$^{\circ}$ which seems to be limited by the geometry of planar samples. This is in agreement with observations for pure CBC7CB (Figure 4b in\cite{2017DenaLC}).

Since in a conventional nematic phase the molecular director is parallel to the rubbing direction (i.e. $\alpha=0$), one could expect existence of materials with a small ($<$22.5$^{\circ}$) initial value of $\alpha$ in the $N_{TB}$ phase. If the average direction of the nanometer-scale helix axis follows the trend defined by the angle $\alpha$, then, in such materials, an increase in this angle towards its saturation value will cause sign inversion in the first harmonic of the electro-optical response. (See Section 2.2 of Ref.\cite{2013SPIE} for details of the experiment).

Though the sign inversion is not detected in the mixture of CBC7CB with 5CB used for SEM imaging presented above, this hypothesis is confirmed for at least two materials of the di-fluoroterphenyl dimer series (Figure 4, inset) with m = 5, n = 7 and m = 5, n = 9 (the compounds DTC5C7 and DTC5C9). When a planar cell is positioned between the crossed polarizers with the rubbing direction at 22.5$^{\circ}$ to the polarizer direction, the optical response at the fundamental frequency (1\emph{f}) of the applied field is proportional to $sin(4(22.5^{\circ}\pm\alpha))$ and its second harmonic (2\emph{f}) is proportional to $cos(4(22.5^{\circ}\pm\alpha))$ (see supplementary information). Therefore, at some point close to the onset of self-deformation striped pattern, the first harmonic will cross the zero value. This sign reversal is clearly visible in the data set presented in Figure 4. Therefore the initial value of $\alpha$ at the transition to the $N_{TB}$ phase on cooling is below 22.5$^{\circ}$. This is in perfect agreement with the values for the director tilt obtained by NMR (Figure 11 in\cite{2016EmsleyPCCP}). The second harmonic signals generated by the change in $\alpha$ in the bands with opposite tilt will be canceled out, thus the 2\emph{f} signal observed is mainly generated by the change in birefringence. This indicates that not only the in-plane position of the optical axis, but also the birefringence is varying with the applied field, i.e. the axis of the nano-scale helix may significantly deviate from the plane of the cell.

The absence of sign reversal of the response in the 20/80\% w/w mixture of 5CB and CB-C7-CB confirms that the value of $\alpha$ stays between 22.5$^{\circ}$ and 45$^{\circ}$ over the entire $N_{TB}$ temperature range, as seen in SEM images.

We note that the handedness of the nano-scale helix must be the same over the area of measurements ($\varnothing$2 mm) covering multiple periods of the self-deformation striped pattern. Otherwise the first harmonic of the signal will be averaged over the pattern to zero.

In summary, we  demonstrate that adding a photopolymerizeable component to a LC material provides an effective tool for detailed investigations of its structure. When combined with SEM imaging, details of the hierarchy of patterns in the $N_{TB}$ phase scaling from 20 nm to several $\mu$m and larger are revealed.

We rule out the presence of the "rope-like" micrometre-scale helix in favour of the "fish-bone" structure. The structure is formed by a more dense irregular helix-like pattern(s) of $\sim$100 nm size. Those, in turn, are formed by $\sim$10 times smaller helical arrangements of molecules reported earlier\cite{2013LavrNatComm,2013ClarkPNAS,ClarkDH,2017StevensonPCCP}.

We also note that the angle between the rubbing direction and the average filament axis is close to 40-45$^{\circ}$ in most cases. This finding has led us to propose the existence of materials with sign reversal of the electro-optic response. An example of such a material is given.

Apart from the clear potential in revealing of complex liquid crystalline structures, the photo polymerization technique in combination with the unique properties of the twist-bend phase will open a novel platform for applications of micro- and nano-technology of advanced functional materials.

\begin{suppinfo}
Additional high-resolution images and mathematical details are available in supplementary information.
\end{suppinfo}

\begin{acknowledgement}
The work as supported by 13/US/I12866 from the Science Foundation Ireland as part of the US-Ireland Research and Development Partnership program jointly administered with the United States National Science Foundation under grant number NSF-DMR-1410649. Financial support for the Hull group was from FP7 EU Grant No. 216025 and the EPSRC project EP/M015726.
\end{acknowledgement}

\end{document}